\documentclass[10pt,twocolumn,twoside]{pnas-new}

\usepackage[doi=false,isbn=false,url=false,eprint=false,defernumbers=true, style=nature, citestyle=numeric-comp, backend=bibtex, sorting=none]{biblatex}
\usepackage{csquotes}
\makeatletter
\AtEveryCitekey{%
  \ifcsundef{blx@entry@refsegment@\the\c@refsection @\thefield{entrykey}}
    {\csnumgdef{blx@entry@refsegment@\the\c@refsection @\thefield{entrykey}}{\the\c@refsegment}}
    {}}
\defbibcheck{onlynew}{%
  \ifnumless{0\csuse{blx@entry@refsegment@\the\c@refsection @\thefield{entrykey}}}{\the\c@refsegment}
    {\skipentry}
    {}}
\makeatother
\AtEveryBibitem{%
  \clearfield{note}%
}
\usepackage{jabbrv}
\AtBeginBibliography{\scriptsize}
  
\usepackage[version=3]{mhchem}
\usepackage[utf8]{inputenc}
\usepackage[T1]{fontenc}
\usepackage{textcomp}
\usepackage{gensymb}
\usepackage{lipsum} 
\usepackage{soul} 
\templatetype{pnasresearcharticle} 

\title{X-ray nanotomography reveals formation of single diamonds by block copolymer self-assembly}

\author[a]{K.\,Djeghdi}
\author[b,c]{D.\,Karpov}
\author[a,d]{S.\,N.\,Abdollahi}
\author[a]{K.\,Godlewska} 
\author[b]{M.\,Holler}
\author[e]{C.\,Donnelly}
\author[f,g]{T.\,Yuasa} 
\author[f,h]{H.\,Sai}
\author[f,i]{U.\,B.\,Wiesner}
\author[a]{U.\,Steiner}
\author[a,j]{B. D.\,Wilts}
\author[k]{M.\,Musya}
\author[k,l,m,n,o]{S.\,Fukami}
\author[k,l,m,n]{H.\,Ohno}
\author[b]{A.\,Diaz}
\author[k,l]{J.\,Llandro}
\author[a,*]{I. Gunkel}

\affil[a]{Adolphe Merkle Institute, University of Fribourg, Chemin des Verdiers 4, 1700 Fribourg, Switzerland}
\affil[b]{Paul Scherrer Institute, Forschungsstr.\,111, 5232 Villigen-PSI, Switzerland}
\affil[c]{Current affiliation: European Synchrotron Radiation Facility, 71 Av. des Martyrs, 38000 Grenoble, France}
\affil[d]{Current affiliation: Department of Chemistry, University of Basel, Mattenstr.\,24a, BPR-1096, 4058 Basel, Switzerland}
\affil[e]{Max Planck Institute for Chemical Physics of Solids, N\"othnitzer Str. 40, 01187 Dresden, Germany}
\affil[f]{Department of Materials Science and Engineering, Cornell University, Ithaca, NY 14853, USA}
\affil[g]{Current affiliation: Yokkaichi Research Center, JSR Corporation, Yokkaichi, Mie 510-8552, Japan}
\affil[h]{Current affiliation: Simpson Querrey Institute for Bionanotechnology, Northwestern University, Evanston, IL 60208, USA}
\affil[i]{Kavli Institute at Cornell for Nanoscale Science, Cornell University, Ithaca, NY 14853, USA}
\affil[j]{Department for Chemistry and Physics of Materials, University of Salzburg, Jakob-Haringer-Str.\,2a, 5020 Salzburg, Austria}
\affil[k]{Laboratory for Nanoelectronics and Spintronics, Research Institute of Electrical Communication, Tohoku University, 2-1-1 Katahira, Aoba-ku, Sendai 980-8577, Japan}
\affil[l]{Center for Science and Innovation in Spintronics, Tohoku University, 2-1-1 Katahira, Aoba-ku, Sendai 980-8577, Japan}
\affil[m]{Center for Innovative Integrated Electronic Systems, Tohoku University, 468-1 Aramaki Aza Aoba, Aoba-ku, Sendai 980-0845 Japan}
\affil[n]{WPI Advanced Institute for Materials Research, Tohoku University, 2-1-1 Katahira, Aoba-ku, Sendai 980-8577, Japan}
\affil[o]{Inamori Research Institute for Science, Kyoto 600-8411, Japan}

\leadauthor{Djeghdi} 

\correspondingauthor{\textsuperscript{*}To whom correspondence should be addressed. E-mail: ilja.gunkel@unifr.ch}


\begin{abstract}
Block copolymers are recognised as a valuable platform for creating nanostructured materials with unique properties. Morphologies formed by block copolymer self-assembly can be transferred into a wide range of inorganic materials, enabling applications including energy storage\supercite{li_self_2020} and metamaterials\supercite{alvarez_block_2021}. However, imaging of the underlying, often complex, nanostructures in large volumes has remained a challenge, limiting progress in materials development. Taking advantage of recent advances in X-ray nanotomography\supercite{holler_high-resolution_2017,michelson_three-dimensional_2022}, we non-invasively imaged exceptionally large volumes of nanostructured soft materials at high resolution, revealing a single diamond morphology in a triblock terpolymer composite network. This morphology, which is ubiquitous in nature, has so far remained elusive in block copolymers, despite its potential to create materials with large photonic bandgaps\supercite{maldovan_diamond_2004}. The discovery was made possible by the precise analysis of distortions in a large volume of the self-assembled diamond network, which are difficult to unambiguously assess using traditional characterisation tools. We anticipate that high-resolution X-ray nanotomography, which allows imaging of much larger sample volumes than electron-based tomography, will become a powerful tool for the quantitative analysis of complex nanostructures and that structures such as the triblock terpolymer-directed single diamond will enable the generation of advanced multicomponent composites with hitherto unknown property profiles.
\end{abstract}

\dates{This manuscript was compiled on \today}

\begin{document}
\newrefsegment
\maketitle
\thispagestyle{firststyle}
\ifthenelse{\boolean{shortarticle}}{\ifthenelse{\boolean{singlecolumn}}{\abscontentformatted}{\abscontent}}{}


X-ray tomography allows non-destructive 3D imaging of large sample volumes due to the large penetration depth of X-rays. Originally used mainly for biological samples\supercite{shapiro_biological_2005,jiang_quantitative_2010}, it is now being extended to materials science, including lithium-ion batteries\supercite{wood_x-ray_2018,sadd_investigating_2023} with recent improvements in resolution now allowing nanoscale imaging\supercite{sakdinawat_nanoscale_2010,withers_x-ray_2021}. X-ray nanotomography\supercite{dierolf_ptychographic_2010,pfeiffer_x-ray_2018,shapiro_ultrahigh_2020}, particularly that based on X-ray ptychography, has recently enabled 3D imaging of nanostructures in integrated circuits\supercite{holler_high-resolution_2017} and nanoparticle arrays\supercite{michelson_three-dimensional_2022}. Despite the enormous utility of this emerging technique for the characterisation of nanostructured materials, its application to self-assembled nanostructures\supercite{wilts_evolutionary_2018} remains largely underrepresented. One area where X-ray nanotomography can be particularly useful is in the study of nanostructured materials generated by the self-assembly of block copolymers, especially those with 3D network structures. Their continuous pathways in all three spatial directions are important for applications involving transport of electrons\supercite{robbins_block_2016} or ions\supercite{sutton_surface_2019}, while their distinct morphologies produce metamaterials with emerging properties\supercite{hur_three-dimensionally_2011,kilchoer_strong_2020} whose performance is closely linked to structural details of the underlying network\supercite{dolan_metasurfaces_2019}. 

However, the identification of networks formed by the self-assembly of block copolymers remains challenging. Unlike their natural counterparts, for example in butterflies\supercite{michielsen_gyroid_2008}, they often exhibit unique symmetry-breaking distortions\supercite{feng_seeing_2019} that are difficult to assess using conventional characterisation techniques such as X-ray scattering\supercite{reddy_block_2021}. The situation is exacerbated by the transition from diblock copolymer to triblock terpolymer-derived materials, as reported herein, which exhibit even more complex phase behaviour, including multiple network structures\supercite{meuler_ordered_2009}. Here we show that X-ray nanotomography allows the unambiguous assignment of a rarely observed network structure in self-assembled synthetic materials. 

\begin{figure*}[p!]
\centering
\includegraphics[width=0.95\linewidth]{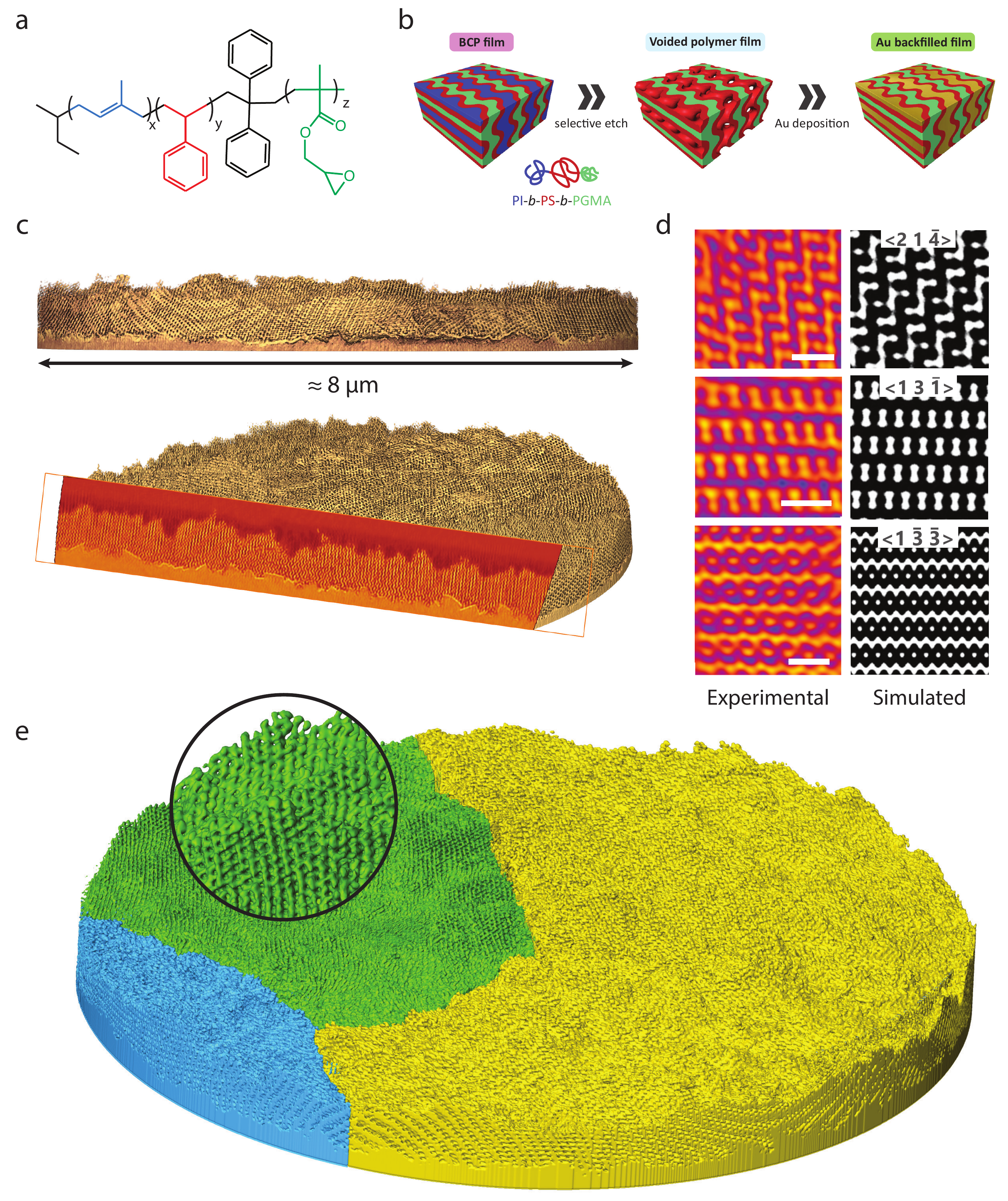}
\caption{\textbf{Replication of a self-assembled polymer network and its volumetric reconstruction.} 
a) Chemical structure of polyisoprene-\textit{block}-polystyrene-\textit{block}-poly(glycidylmethacrylate) (PI-\textit{b}-PS-\textit{b}-PGMA, ISG), where PI is shown in blue, PS in red, and PGMA in green.
b) Scheme of sample preparation: from an ISG terpolymer film prepared by solvent vapour annealing, the PI phase is selectively removed to form a voided film, which is backfilled with gold by electrodeposition. PI is shown in blue, PS in red, PGMA in green, and gold in yellow. 
c) 3D reconstruction of the gold network determined by PXCT and exemplary cross-section. 
d) Comparisons between cross-sections extracted from the reconstructed volume obtained by PXCT (left) and cross-sections of the single diamond (SD) according to the level-set model in eq.~\ref{eq:level-set-diamond} (right). The Miller indices of the SD level-set are written as $\langle hkl\rangle$. Scale bars: 100\,nm. 
e) Full volume reconstruction of the sample ($d \approx 8$\,\textmu m) with grains highlighted in different colours.}
\label{fig1:replication_volumes}
\end{figure*}


The experimental system consists of the previously unreported poly\-iso\-prene-\emph{b}-poly\-sty\-rene-\emph{b}-poly\-(gly\-ci\-dyl\-meth\-ac\-ry\-late) (PI-\emph{b}-PS-\emph{b}-PGMA, ISG) triblock terpolymer, the chemical structure of which is shown in Figure\,\ref{fig1:replication_volumes}a. ISG was synthesised by sequential anionic polymerisation according to established procedures\supercite{bailey_morphological_2001,hadjichristidis_linear_2005} described in the Methods. The asymmetric composition of the ISG terpolymer, with volume fractions of $f_\mathrm{PI} = 0.29$, $f_\mathrm{PS} = 0.52$, and $f_\mathrm{PGMA} = 0.19$, allows curved interfaces between the phases formed by the different blocks, characteristic of 3D network structures, while its molar mass of 67.4\,kg/mol is sufficiently high to induce self-assembly.\supercite{meuler_ordered_2009} Large grains of 3D block copolymer networks are typically formed by slow solvent casting, often over several days.\supercite{feng_seeing_2019,chang_mesoscale_2021} A similar protocol was used here to produce a well-ordered nanostructured network in an ISG terpolymer film by controlled swelling in the vapour of the organic solvent tetrahydrofuran (THF), followed by slow drying over a period of 44\,h. The dried film was voided by selective removal of the PI phase. The resulting nanoporous polymer template was then backfilled with gold by electrodeposition to replicate the network structure created by the PI phase\supercite{vignolini_3d_2012} (Figure\,\ref{fig1:replication_volumes}b; details of the sample preparation are given in the Methods). 

\begin{figure*}[tbp]
\centering
\includegraphics[width=0.95\linewidth]{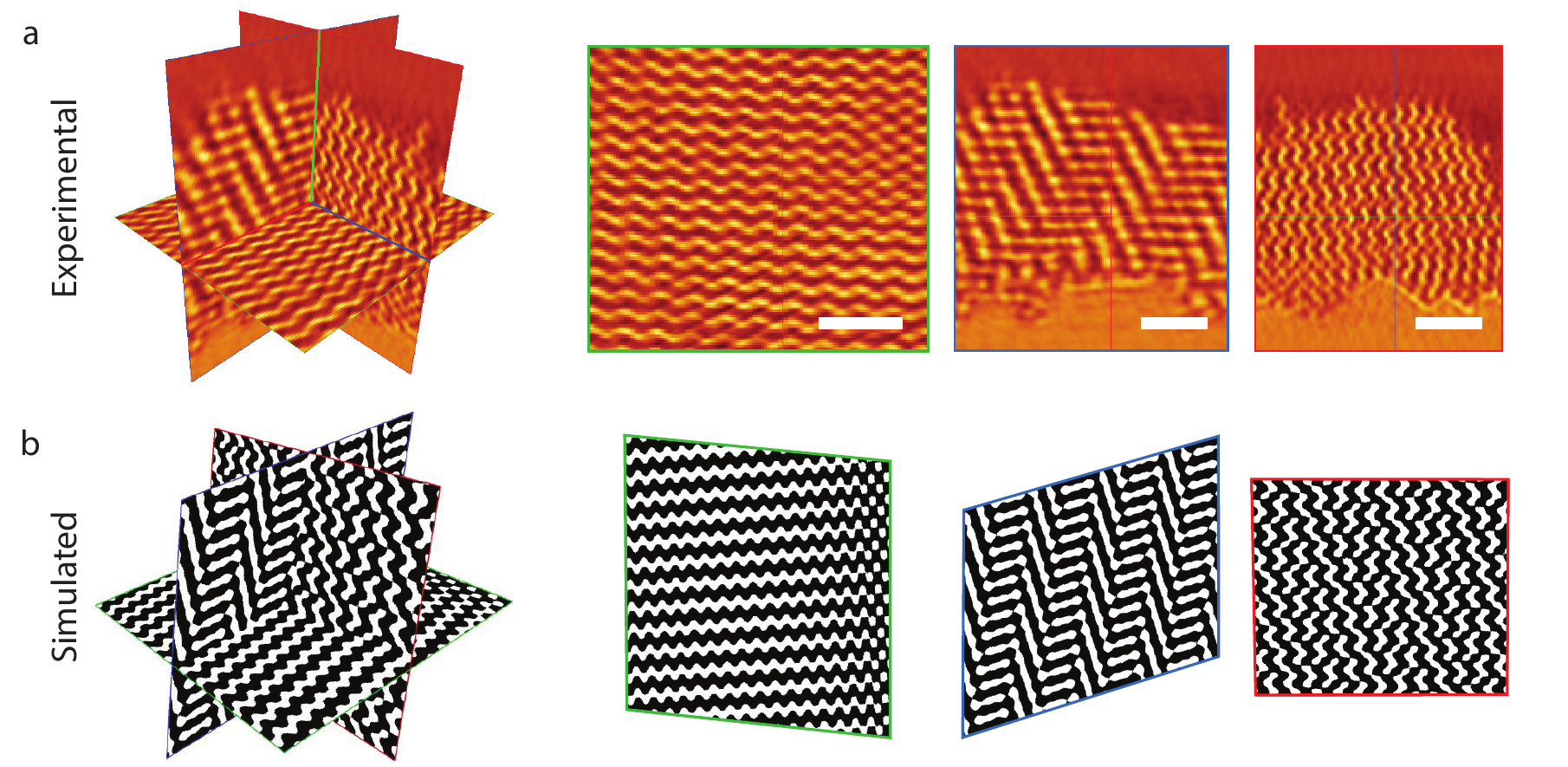}
\caption{\textbf{Experimental and simulated orthoslices.} a) Sub-volume of the tomogram with orthogonal slices shown on the left and the corresponding cross-sections with colour-matched frames shown on the right. Scale bars: 200\,nm. 
b) Cross-sections of an SD level set computed from eq.\ \ref{eq:level-set-diamond}, matching the experimental cross-sections of a). The angles between the computed slices are $\alpha_{\mathrm{gb}}=92\degree$, $\alpha_{\mathrm{gr}}=107\degree $, $\alpha_{\mathrm{rb}}= 97\degree$, indicating that the experimental network is distorted compared to the ideal SD in the level set model. The green outlined cross-section, corresponding to a substrate-parallel orientation, is close to the (110) plane of the SD.
}
\label{fig2:CS}
\end{figure*}

For the unambiguous morphological assignment of 3D network structures, tomographic imaging and 3D reconstruction of the structure are essential. Established 3D imaging using TEM tomography typically produces volumes of 10s to 100s of unit cells, allowing quantitative structural analysis\supercite{li_linking_2014}, while more recent FIB-SEM tomography extends this to 1000s of unit cells\supercite{feng_seeing_2019,reddy_block_2021}. The ptychographic X-ray computed tomography (PXCT)\supercite{dierolf_ptychographic_2010,holler_high-resolution_2017,michelson_three-dimensional_2022} reported here allows imaging of about 70,000 unit cells of a 3D network structure (as detailed below) with an estimated resolution of approximately 11\,nm, \textit{i.e.} substantially larger sample volumes than with any previous technique. From a cylindrical pillar sample with a diameter of 8\,\textmu m, the 3D gold network within a PS polymer matrix was reconstructed with a voxel size of 6.04\,nm, resulting in the tomogram shown in Fig.\,\ref{fig1:replication_volumes}c,e. The PXCT experiment and the pillar sample preparation are described in detail in the Methods. 

Morphological analysis of the network is greatly facilitated by the ability to extract a virtually unlimited number of cross-sections from the tomogram, which can be compared with cross-sections calculated from model structures. Several unique patterns were observed in the cross-sections extracted from the reconstructed volume of the 3D gold network at various orientations. Despite some angular distortion, a matching plane was found in the single diamond (SD) network for each of these cross-sections (Figure\, \ref{fig1:replication_volumes}d and Extended Data Fig.\,\ref{figSI:add_CS}). The quality and number of matching cross-sections found between the experimental and calculated data allow confident assignment of the network to the SD. For the calculation of the planes in an ideal SD -- a triply periodic minimal surface  with space group $Fd\bar{3}m$ (Q$^{227}$) -- its isosurface was approximated by the following level-set equation\supercite{michielsen_gyroid_2008}:
\begin{equation}
  \cos(Z)\sin(X+Y)+\sin(Z)\cos(X-Y)<t,
  \label{eq:level-set-diamond}
\end{equation}
where $t=2.4\left(f-0.5\right)$ is a parameter related to the fill fraction $f$ of the solid network phase. A volumetric SD is then defined by filling the channels bounded by one side of the isosurface (the other side being the matrix) while its skeleton is the network passing through the middle of each strut. 

For the ISG terpolymer network, this means that a PS matrix separates SDs of PI and PGMA, forming what is known as an alternating diamond \textsuperscript{A}D. Figure\,\ref{fig1:replication_volumes}b (BCP film) shows an ISG \textsuperscript{A}D, plotted from the level-set equation (eq.\ \ref{eq:level-set-diamond}) with $f_\mathrm{PI} = 0.29$, $f_\mathrm{PS} = 0.52$, and $f_\mathrm{PGMA} = 0.19$. In general, the \textsuperscript{A}D is characterised by the same $Fd\bar{3}m$ space group (Q$^{227}$) as the SD, whereas the double diamond (DD), for which both networks are made of the same material, is described by the $Pn\bar{3}m$ space group (Q$^{224}$). This distinction is crucial for self-assembled structures: since the two networks are made of different materials, only an \textsuperscript{A}D can serve as a template for an SD that can be accessed by the described gold replication protocol (see Figure\,\ref{fig1:replication_volumes}b, Au backfilled film). 

In a more refined and quantitative analysis, three pairwise orthogonal slices extracted from the tomogram (yellow grain in Fig.\,\ref{fig1:replication_volumes}e), one parallel and two perpendicular to the substrate (Fig.\,\ref{fig2:CS}a), were matched to pairwise (approximately) orthogonal cross-sections of a level-set generated SD (Fig.\,\ref{fig2:CS}b). This simultaneous agreement between experimental and simulated slices in all three spatial directions not only proves the morphological assignment of an SD, but also allows the orientation of the network to be determined and its distortion to be quantified. Based on the matching orthogonal slices, a substrate normal of the SD of $\vec{n}=(0.71, 0.70, 0.03)$ was determined, approximating a $\langle110\rangle$ out-of-plane orientation, \textit{i.e.} a substrate-parallel orientation of the (110) planes. The two micrometre-wide grains reconstructed from the micropillar sample (yellow and green grains in Fig.\,\ref{fig1:replication_volumes}e) both show a uniform orientation of their $(110)$ planes across the thickness of the extracted sample (Extended Data Fig.\,\ref{figSI:vertstack}). We imagine that such uniformity in the orientation of the sample is valuable, for example, in the interaction with light.    

Following previous arguments\supercite{hashimoto_identification_2007}, we rationalise the observed orientation of the lattice underlying the self-assembled ISG terpolymer network by comparing the composition of the $(110)$ and other distinct planes of the \textsuperscript{A}D morphology (Extended Data Fig.\,\ref{figSI:ff_variation}). The composition of the ${110}$ planes differs from the other planes in that: i) it has the largest variation in the volume fraction of the matrix phase, and ii) it has the overall highest value of the maximum volume fraction of the matrix phase (74\%), which represents a significant deviation from its bulk volume average of 52\% ($=f_\mathrm{PS}$). This maximum is reached when the other two phases (PI and PGMA) form centred rectangular lattices shifted by half a unit cell (Extended Data Fig.\,\ref{figSI:ff_variation}a). Preferred lattice orientations are commonly observed for other cubic morphologies in self-assembled block copolymer films. The double gyroid and the alternating gyroid, for example, favour a substrate-parallel orientation of their $(211)$\supercite{lee_structural_2005}
and $(110)$
\supercite{suzuki_the_2000} planes, respectively. Analogous to the alternating diamond studied here, these are the planes with the greatest compositional variation. Preferred orientations also extend to non-cubic morphologies, such as block copolymer lamellae and cylinders, which typically orient in the plane of the film\supercite{fasolka_block_2001,hamley_ordering_2009}. While this behaviour is commonly attributed to the preferential wetting of one of the blocks at the film interfaces, the orientation of a block copolymer morphology also depends on the annealing kinetics and the solvent used\supercite{paradiso_block_2014}.  

\begin{figure}[tb!]
\centering
\includegraphics[width=0.95\linewidth]{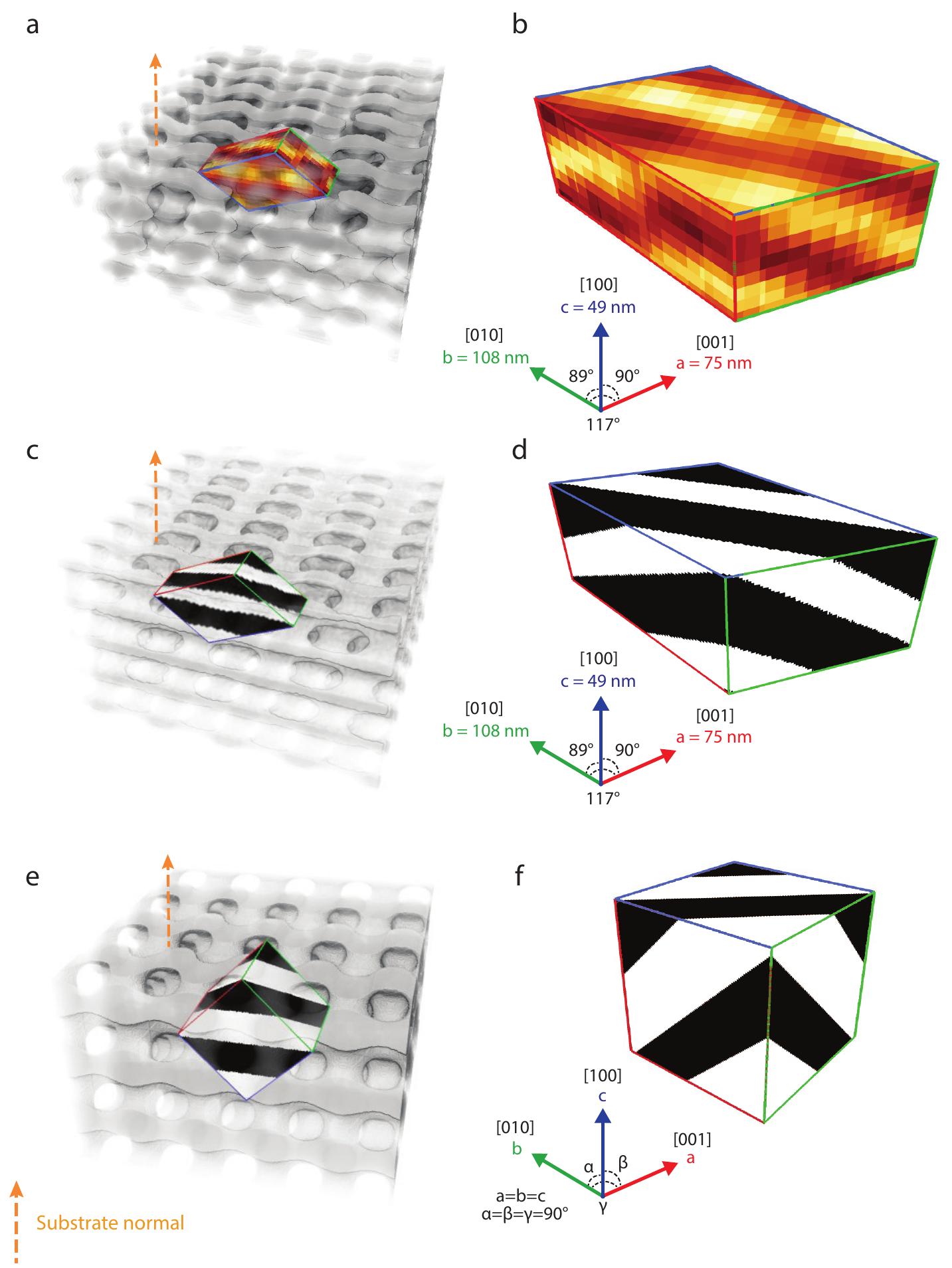}
\caption{\textbf{Unit cell distortion.} a) Conventional unit cell within a translucent volume rendering of the reconstructed network, with the substrate-normal facing up as indicated by the orange arrow.
b) The conventional unit cell extracted from the tomogram along with the determined unit cell parameters ($a,b,c,\alpha, \beta, \gamma$). The stripe patterns seen on the unit cell surface are characteristic of the $\{100\}$ family of planes. The plane with the blue outline is parallel to the sample normal. 
c) Conventional unit cell in a translucent volume rendering of the DSD with the $\left[110\right]$ direction pointing upwards, which corresponds to the orientation of the experimental data.
d) DSD unit cell resulting from the affine transformation of a cubic unit cell according to eq.\ \ref{eq:matrix_affine_transform}.
e) Conventional cubic unit cell in a translucent volume rendering of an SD with the $\left[110\right]$ direction pointing upwards.
f) Conventional cubic unit cell of an SD level set plotted from eq.\ \ref{eq:level-set-diamond} with a fill fraction $f=0.4$.} 
\label{fig3:unitcell}
\end{figure}

Based on the assignment of the orientation, the \{100\} family of planes was located in the reconstructed volume to extract the experimental conventional unit cell (Fig.\,\ref{fig3:unitcell}). This gives unit cell parameters of 75\,nm, 108\,nm, and 49\,nm, and angles between the $\left[100\right]$, $\left[010\right]$ and $\left[001\right]$ directions of $\alpha=89$\textdegree, $\beta=90$\textdegree, and $\gamma=117$\textdegree, approximating a monoclinic unit cell. Its evident angular deformation is most pronounced between the two planes separated by $\gamma$. 
Note that the structure remained stable during the X-ray exposure (Extended Data Fig.\,\ref{figSI:PXCT}). The observed deformation is consistent with the vertical shrinkage that self-assembled morphologies within films typically undergo during drying (Fig.\,\ref{fig3:unitcell}a). Shrinkage during solvent processing has previously been shown to induce significant lattice distortions in cubic gyroids\supercite{dolan_controlling_2018,jo2021symmetry}. Since the unit cell obtained here also differs from a cubic unit cell (Fig.\,\ref{fig3:unitcell}f), a deformed single diamond (DSD) was generated by modifying the SD level-set model to match the experimental unit cell parameters. This was achieved by scaling and shearing an SD with all lattice parameters equal to about 71\,nm, which corresponds to a volume-conserving 3D affine transformation represented by the following matrix: 
\begin{equation}
 T= \begin{bmatrix}
S_x & -0.5095 S_x & 0 \\
0 & S_y & 0 \\
0 & -0.018 S_z & S_z \\
 \end{bmatrix},
\label{eq:matrix_affine_transform} 
\end{equation} 
where $S_x\approx0.94$, $S_y\approx1.53$ and $S_z\approx0.72$ are the scaling factors. The resulting DSD is shown in Fig.\,\ref{fig3:unitcell}c,d and is in excellent agreement with the experimentally determined unit cell shown in Fig.\,\ref{fig3:unitcell}a,b. The model, therefore, suggests that the observed monoclinic distortion of the unit cell can be described by affine transformations of an ideal cubic SD.

\begin{figure*}[tbp]
\centering
\includegraphics[width=\linewidth]{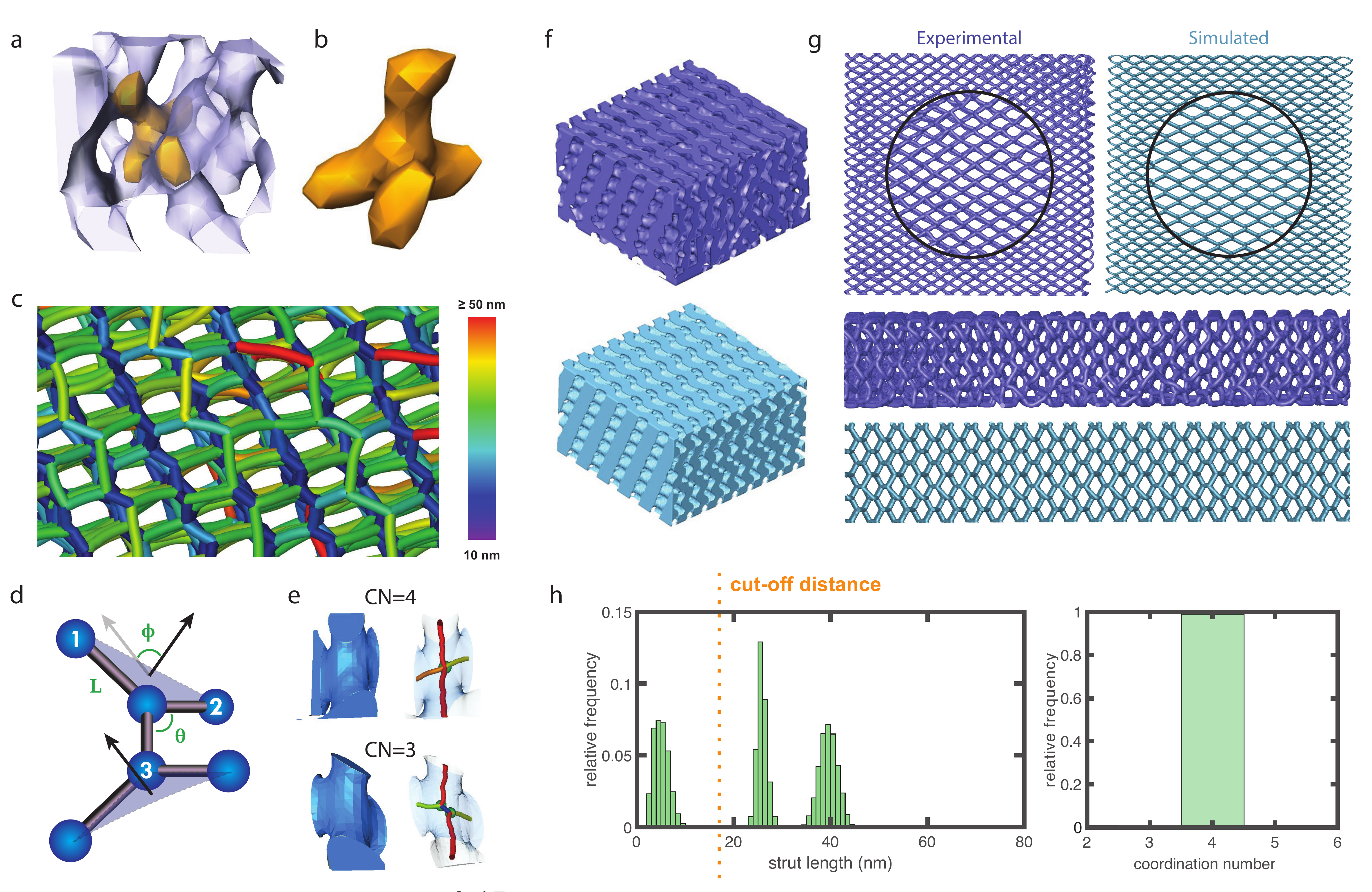}
\caption{\textbf{Skeletal analysis of the ptychography data.} 
a) Subvolume ($V\approx(145\,\mathrm{nm})^3$) of the ptychography dataset with a single node highlighted in yellow. 
b) Extracted single node. 
c) Skeleton of a subvolume of the ptychography dataset colour-coded with the strut length. 
The short blue struts having a common upward direction are likely caused by shrinkage as the film dries.  
d) Definition of strut length $L$, bond angle $\theta$, coordination number CN, and dihedral angle $\varphi$.
e) Examples of 3-connected and 4-connected struts in a model DSD.
f) Two equivalent subvolumes extracted from the experimental dataset (dark blue) and the model DSD (light blue).
g) Comparison of top- and side-view skeletons of subvolumes ($V=430\times430\times260$\,nm\textsuperscript{3}) extracted from the experimental dataset (dark blue) and the model DSD (light blue).
h) Strut length (left) and coordination number distributions (right) computed for a model DSD with a resolution of 1\,nm and a minimum strut length of 18\,nm.}
\label{fig4:node_skelet_3Dstat}
\end{figure*}

The exceptionally large volume of the full reconstructed network with approximately 70,000 unit cells combined with the small voxel size of 6.04\,nm provides an excellent opportunity for statistically meaningful structural analysis down to individual network nodes. This includes the network structure both within a single grain and close to a grain boundary (yellow grain in Fig.\,\ref{fig1:replication_volumes}e), which is discussed in a separate manuscript\supercite{karpov2023highresolution}. A subvolume of the full reconstructed data and a representative fourfold connected node are shown in Fig.\,\ref{fig4:node_skelet_3Dstat}a,b. Skeletonising such a 3D dataset, \textit{i.e.}\ replacing volumetric struts connecting two vertices with their medial axis, generates a ball-and-stick model from which structural parameters such as the strut length $L$, bond angle $\theta$, dihedral angle $\phi$ and coordination number CN can be extracted (Fig.\,\ref{fig4:node_skelet_3Dstat}d,e). For the experimental dataset and the computed DSD, the investigated volumes are $V \approx 1.0 \times 1.5 \times 0.3\,\mathrm{\micro m}^3$, containing approximately 10,000 nodes. The DSD model agrees well with the experimental data set, both qualitatively (Fig.\,\ref{fig4:node_skelet_3Dstat}f,g) and in the associated quantitative structural analysis (Extended Data Fig.\,\ref{figSI:vox_stat_ptycho}).

Given this good agreement, the structural parameters were assessed using the skeleton of a high-resolution DSD model (voxel size 1\,nm), which enables a refined structural analysis. The skeleton of this DSD model exhibits a trimodal distribution of strut lengths, centred around 5, 25 and 40\,nm (Fig.\,\ref{fig4:node_skelet_3Dstat}h). The skeleton of the experimental data (Fig.\,\ref{fig4:node_skelet_3Dstat}c) features similarly short (blue) struts in the direction approximately orthogonal to the substrate. However, a closer analysis of the skeletonisation (Fig.\,\ref{fig4:node_skelet_3Dstat}e and Extended Data Fig.\,\ref{figSI:vox_node}) reveals that these struts result from resolution-dependent skeletonisation artefacts and are therefore artificial, unlike the bridging struts observed in double gyroid samples using a high-resolution TEM tilt-series \supercite{miyata2022dislocation}. Note also that the voxel size should ideally be at most one tenth of the size of the smallest feature in a sample in order to obtain a faithful volumetric reconstruction\supercite{holzer_three_2004}. Imposing a minimal strut length of 18\,nm removes these unphysical struts by merging two closely neighbouring nodes into one, which yields a perfect fourfold connectivity (Fig.\,\ref{fig4:node_skelet_3Dstat}h and Extended Data Fig.\,\ref{figSI:vox_stat_DSD}) characteristic of the diamond structure (Extended Data Fig.\,\ref{figSI:diamond}a,c).  
The distribution of bond angles shows one peak centred around $\theta=80\degree$\ and three additional peaks between $120-140\degree$  ($N\approx30.000$), while the distribution of the dihedral angles displays main peaks at approximately $\{\pm 180\degree, 0\degree\}$ ($N\approx 25.000$) (Extended Data Fig.\,\ref{figSI:vox_stat_DSD}). Given the unit cell distortion described above, which results from shrinkage of the terpolymer film during drying, these values differ from the values expected for a level-set cubic SD shown in Extended Data Fig.\,\ref{figSI:diamond} ($\mathrm{CN}=4$, $\theta=109.5\degree$, $\varphi=\pm 60\degree, \pm 180\degree$). The experimental structure is, however, closely related to an SD (see Extended Data Fig.\,\ref{figSI:diamond}a,b), in particular retaining its fourfold connectivity.


While the SD is commonly observed in the exoskeletons of beetles and weevils\supercite{han_overview_2018}, it has only recently been achieved synthetically, for example in liquid crystals\supercite{zeng_self-assembled_2019} and colloids\supercite{he_colloidal_2020}.
Our observation of an alternating diamond (consisting of two shifted SDs) in a terpolymer may also be surprising. Indeed, from a thermodynamic standpoint, its formation is predicted only over a very limited region in phase space\supercite{qin_phase_2010}. The scarcity of self-assembled BCP diamonds, especially the alternating diamond observed here, is associated with their higher free energy compared to gyroid networks. This is because three-connected gyroid nodes exert less entropically penalised stretching on the polymer chains than the four-connected diamond nodes\supercite{matsen_stable_1994,matsen_origins_1996}. This unfavourable chain stretching in copolymers can be relieved if the nodes are sufficiently filled with other molecules\supercite{matsen_phase_1995}. Following this idea, block copolymer diamonds were stabilised, for example, by adding homopolymer\supercite{takagi_ordered-bicontinuous-double-diamond_2015}, small molecules\supercite{sheng_selfassembly_2021}, or by blending copolymers of different chain lengths\supercite{asai_tricontinuous_2017,takagi_bicontinuous_2019}. 
While in the present study solvent molecules were added to the ISG terpolymer during processing, the final structure is dry. The diamond produced therefore consists only of terpolymer. Skeletal analysis of the high-resolution tomogram revealed a distortion of the diamond network and its nodes. However, the effect of such elongated nodes on the overall free energy of the diamond and the degree of chain stretching remains unclear. Alternatively, the diamond may have been stabilised in the presence of the solvent, and then kinetically trapped in the dry film, as previously seen in other network structures\supercite{chang_mesoscale_2021}. These open questions could be addressed using computational approaches such as self-consistent field theory (SCFT). 

Our observations have been made possible by recent advances in X-ray nanotomography, which now allows for high-resolution imaging of large sample volumes. Analysis of up to 70,000 unit cells in a self-assembled triblock terpolymer network with sub-15 nm resolution enabled identification of the elusive single diamond morphology. Such high-resolution, real-space 3D structural information allows the unambiguous assignment of such complex network morphologies. It also allows quantification of relevant structural distortions, which often complicate structural analysis based solely on scattering experiments, and together with the study of mesoscale defects and grain boundaries\supercite{karpov2023highresolution}, represents an emerging area in soft matter crystal analysis\supercite{feng_seeing_2019}. In future studies, the non-invasive nature of X-ray nanotomography may further allow such detailed structural information to be directly coupled with subsequent material property investigations. This, in turn, may provide opportunities to establish structure-property relationships at unprecedented levels of structural detail. This is particularly relevant for nanostructured multicomponent materials such as the polymer-metal composites studied here, where structural details are expected to have substantial effects, \textit{e.g.}, on photonic/plasmonic material properties.\supercite{dolan_metasurfaces_2019} 

 \vspace{0.5cm}

\newpage
\section*{Methods}
\newrefsegment

\subsection*{Triblock terpolymer synthesis}
Standard Schlenk line techniques were used throughout the synthesis of the poly\-iso\-prene-\emph{block}-poly\-sty\-rene-\emph{block}-poly\-(gly\-ci\-dyl\-meth\-ac\-ry\-late) (PI-\emph{b}-PS-\emph{b}-PGMA, ISG) triblock terpolymer. Diphenylethylene-end-capped PI-\textit{b}-PS (IS-DPE) was synthesised from isoprene (99 \%, Sigma-Aldrich), styrene ($\geq$99 \%, Sigma-Aldrich) and 1,1-diphenylethylene (Sigma-Aldrich) by sequential anionic polymerisation\supercite{creutz_living_1997,li_linking_2014} using \textit{sec}-butyllithium (Sigma-Aldrich) in benzene. The solvent was exchanged to tetrahydrofuran (THF) containing 5 equiv. lithium chloride with respect to the sec-butyllithium by cryogenic-assisted vacuum distillation (anhydrous, inhibitor-free, 99 \%, Sigma-Aldrich). Glycidyl methacrylate (>95 \%, TCI America) monomer was vacuum distilled over calcium hydride and filtered through active alumina (Fisher Scientific) in an inert atmosphere glove box, then vacuum distilled again over calcium hydride\supercite{hild_diblock_1998}. In both distillation processes, the first and the last fractions of the distillate were removed. Glycidyl methacrylate monomer was slowly added to IS-DPE in THF at \textminus78\textdegree C and allowed to react for two hours. The polymer was then quenched with degassed methanol (Macron Chemicals), immediately precipitated in methanol, and then dried under reduced pressure for 48\,h. The total molar mass of the ISG terpolymer was 67.4\,kg/mol, as determined by a combination of gel permeation chromatography of the PI block (GPC, Extended Data Fig.\,\ref{figSI:GPC}) and proton nuclear magnetic resonance spectroscopy of the final ISG terpolymer (\textsuperscript{1}H NMR, Extended Data Fig.\,\ref{figSI:NMR}). The polydispersity index of the final ISG terpolymer determined by GPC was 1.08 and its block volume fractions determined by \textsuperscript{1}H NMR were $f_\mathrm{PI} = 0.29$, $f_\mathrm{PS} = 0.52$, and $f_\mathrm{PGMA} = 0.19$  (Extended Data Fig.\,\ref{figSI:NMR}). 

\subsection*{Preparation of polymer films} The ISG terpolymer film was prepared using fluorine-doped tin oxide (FTO)-coated glass as substrate. Prior to film processing, the FTO-coated glass was etched in Piranha solution and subsequently functionalised by immersion in a 4.3\,mM solution of octyltrichlorosilane (Sigma-Aldrich) in anhydrous cyclohexane (Sigma-Aldrich) for 12\,s. The polymer film on an FTO-coated glass substrate was prepared by spin-coating a 10\,wt\% solution of ISG terpolymer in anhydrous anisole (Sigma-Aldrich) for 60\,s at 1200\,rpm with an acceleration of 500\,rpm/s. Using oxygen plasma cleaned (Diener MRC 100, 100\,W, 2\,min) silicon substrates, the above protocol resulted in approximately 600\,nm thick ISG terpolymer films as determined by thin-film interferometry (not shown). 

\subsection*{Solvent vapour annealing (SVA)} 
The ISG terpolymer film was annealed in a sealed polyether ether ketone (PEEK) chamber
. Two gas lines were connected to the chamber inlet: one pure nitrogen line and one nitrogen line that passed through a solvent reservoir before entering the chamber to enrich the nitrogen carrier gas with solvent vapour. Gas flow in these lines was controlled by digital mass flow controllers (MKS Type MF1) with a maximal flow of 100\,sccm. Adjusting the flow through the nitrogen and solvent vapour lines allows control of the concentration and the composition of the solvent vapour environment in the chamber. A manual mass flow controller connected to the exhaust line was used to adjust the pressure inside the chamber. THF (Sigma-Aldrich) was used as a solvent.

The temperature in the annealing chamber was held constant at 21.3\textdegree\,C, controlled by a Peltier element mounted on a copper plate that in turn served to mount the sample. The temperature of the water bath (ThermoFisher ARCTIC SC150-A10B Refrigerated Circulator) containing the solvent reservoir was fixed at $\approx$ 23.0\textdegree C. The ambient relative humidity was 
between 59\,--\,64$\%$ during the 
SVA 
experiments. The temperature of the room was controlled to $\approx 23.7\,^\circ$C. 

The ISG terpolymer film was annealed in THF vapour to a maximum swelling ratio of $\varphi \approx 2.1$ before being subjected to controlled drying over a period of 44 hours. The swelling ratio $\varphi$, calibrated by measurements of ISG terpolymer films on silicon substrates (not shown), is a dimensionless measure of the relative film thickness and is defined as $\varphi =t/t_0$, where $t$ is the thickness of the swollen film and $t_0$ is the initial thickness of the film after spin coating. 

\subsection*{Templated electrodeposition of gold}
To create gold replicas of the terpolymer gyroids, the PI block of ISG terpolymer films was degraded by exposure to UV light (Fisher Scientific, $\lambda=254$\,nm, 15\,W, 11\,cm distance between the UV lamp and the sample) for 15\,min, and subsequently removed by immersion of the terpolymer film in ethanol for 30\,min. The resulting voided network in the terpolymer films was replicated into gold by electrodeposition using a three-electrode cell with Ag/AgCl with KCl (Metrohm) as the reference electrode, a Pt electrode tip (Metrohm) as the counter electrode, and the FTO-coated glass substrate as the working electrode. A cyclic voltammetry scan between $-0.4$\,V and $-1.15$\,V at a rate of $0.05$\,V/s was used to nucleate gold crystals, while deposition was performed at a constant potential of $-0.756$\,V to subsequently fill the terpolymer template with gold. Electroplating was performed using an AutoLab PGSTAT302N potentiostat (Metrohm) with an Au plating solution (matt ECF 60, Metalor) modified by the addition of 0.5\%\,v/v of a brightener consisting of a 66.7\,mM aqueous solution of \ce{As2O3} (Sigma-Aldrich). KOH was used to adjust the pH to approximately 14.

\subsection*{Micropillar sample preparation with focused ion beams}
To perform tomographic measurements, single-diamond micropillar samples were extracted from the gold-filled polymer template (Figure\,\ref{fig1:replication_volumes}b). The micropillar was shaped by FIB (Focused Ion Beam) processing using a Carl Zeiss Microscopy NVision40 FIB-SEM with Ga ion source. First, a layer of carbon ~1.5 \textmu m thick was deposited on the surface of the gold-filled polymer film in the FIB sample chamber, in order to prevent damage to the single-diamond layer by the ion beam. Next, a 30 kV/13 nA \ce{Ga+} beam excavated a square area of the protection layer and underlying layers to ~12 \textmu m depth to begin defining a pillar-shape. The final shape and diameter (approx. 8\,\textmu m) of the pillar-shaped samples measured in the study was refined by a 30\,kV/3\,nA \ce{Ga+} beam, used to excavate to the final ~15\,\textmu m depth to minimise the possibility of beam damage to the pillar.

After FIB milling, each pillar sample was picked up by a Kleindiek NanoControl NC40 micromanipulator system integrated with the NVision40 FIB/SEM. The micromanipulator was fixed to the top of the pillar’s carbon protection layer by depositing ~0.3\,\textmu m thick carbon patch and the bottom of the pillar was cut from the glass substrate by a 30 kV/150\,pA \ce{Ga+} beam. Then the pillar was picked up and fixed close to the tip of a custom OMNY pin\supercite{holler_omny_2017} (see separate publication for details\supercite{karpov2023highresolution}) by further carbon deposition around its base.

\subsection*{Ptychographic X-ray computed tomography (PXCT)}
\subsubsection*{Instrumentation and data acquisition} 
PXCT measurements were performed at the cSAXS beamline of the Swiss Light Source at the Paul Scherrer Institute in Villigen, Switzerland. In X-ray ptychography, the sample is raster-scanned across a coherent, confined beam, in such a way that neighbouring illuminated areas on the sample partially overlap.\supercite{pfeiffer_x-ray_2018} At each scan position, coherent far-field diffraction patterns are recorded in a transmission geometry. Iterative phase retrieval algorithms are then used to reconstruct the image of the sample with quantitative phase and absorption contrast.\supercite{pfeiffer_x-ray_2018,rodenburg_A_2004} In ptychography, the spatial resolution is not limited by the size of the illumination or the step size of the scan, but by the angular extent of the scattering at which the diffraction patterns can be measured with a sufficient signal-to-noise ratio. In practice, however, mechanical vibrations can limit the resolution, and positioning error motions can cause distortions in the reconstructed image. To suppress such vibrations, we used the flexible tOMography Nano Imaging (flOMNI) end-station\supercite{holler_x-ray_2014,holler_high-resolution_2017} (see also separate publication\supercite{karpov2023highresolution}). This instrument features accurate sample positioning with respect to the beam-defining optics using external laser interferometry combined with sample rotation capabilities for tomography\supercite{holler_error_2015}. This setup ensured distortion-free acquisition of projections of the sample at 2400 equally spaced angular positions ranging from 0 to 180 degrees. The total data acquisition time was approximately 35 hours, including the overhead time due to stage movement in between acquisitions. We estimate that the total dose deposited on the sample during this measurement was approximately $1.0\times10^{10}$ Gy. 

For the PXCT measurements, coherent X-rays with a photon energy of 6.2\,keV were focused through a Fresnel zone plate (FZP) of 170\,\textmu m diameter and 60\,nm outermost zone width, in combination with a central stop and an order-sorting aperture, to define a coherent illumination on the sample with a flux of $3.3\times10^8$\,photons/s. The FZP had locally displaced zones specifically designed to produce an optimal non-uniform illumination for ptychography\supercite{odstrvcil_towards_2019}. The sample was placed at about 0.5 mm\,downstream of the focus, where the beam had a size of about 2\,\textmu m. For the ptychographic scans, we used a combined motion between the sample and the FZP, as described elsewhere\supercite{odstrcil_fast_2019}, to minimise the scanning time. Ptychographic scans were performed following the positions of a Fermat spiral pattern\supercite{huang_optimization_2014} with an average step size of 0.5 \textmu m, covering a field of view of $12.5\times3.7$\,\textmu m$^2$. At each scan position, diffraction patterns were recorded with an acquisition time of 0.1\,s using an in-vacuum 1.5M Eiger detector\supercite{dinapoli_eiger_2011} with a pixel size of $75\times75$\,\textmu m$^2$. The detector was placed at a distance of 2.264\,m from the sample inside an evacuated flight tube to reduce air absorption and scattering background. The ptychographic projections were acquired in a non-sequential angular order according to a binary decomposition scheme\supercite{kaestner_spatiotemporal_2011}. In this way, 8 subtomograms, each with 8 times the angular spacing, were acquired consecutively in such a way that the combination of all of them resulted in a uniform angular space. In this way, the stability of the sample could be assessed during acquisition (see Extended Data Fig.\,\ref{figSI:PXCT}). The dose $D$ imparted on the sample was estimated as $D=\mu/\rho N_0 EN_{\mathrm{p}}$\supercite{howells_an_2009}, where $\mu$ is the linear attenuation coefficient, $\rho$ is the mass density of the sample material, $N_0$ is the flux density incident on the sample in photons per unit area, $E$ is the photon energy, and $N_{\mathrm{p}}$ is the number of projections. For the linear attenuation coefficient we used the value for Au\supercite{henke_x-ray_1993}.

\subsubsection*{Ptychographic and tomographic reconstruction}
Ptychographic reconstructions were performed using the Ptycho Schelves package developed by the X-ray Coherent Scattering group at the Paul Scherrer Institute\supercite{wakonig_ptychoshelves_2020}. For the reconstructions, we selected 1000×1000 detector pixels, resulting in a reconstructed pixel size of 6.04 nm. For each 2D projection, we used 500 iterations of a least-squares maximum-likelihood algorithm with compact set approach\supercite{odstrvcil_iterative_2018}. All phase images acquired at different angles were registered with sub-pixel accuracy using methods developed by the same group\supercite{guizar-sicairos_phase_2011,odstrvcil_alignment_2019}. In addition to registration, these methods allowed an assessment of the sample stability during acquisition, confirming that the high dose delivered to the sample did not cause any significant change in its structure at the achieved resolution (see Extended Data Fig.\,\ref{figSI:PXCT}). After registration, tomographic reconstruction was performed by filter back projection using a Ram Lak filter with a frequency cut-off of 1. A Fourier shell correlation analysis\supercite{vanheel_fourier_2005} of the resulting 3D dataset gave a resolution estimate of approximately 7\,nm, while line profiles across the dataset gave a more conservative resolution estimate of approximately 11\,nm, as described elsewhere\supercite{karpov2023highresolution}. 

\subsection*{3D reconstruction and analysis}
Image stacks were processed using Fiji \supercite{schindelin_fiji_2012} and FEI Avizo\textsuperscript{\texttrademark} for Materials Science 2020.2 software for basic image processing, 3D reconstruction and statistical analysis. A median filter was applied to despeckle and smooth the images. Segmentation was performed in Fiji using the trainable Weka 3D segmentation plug-in. \supercite{arganda-carreras_trainable_2017} Skeletonisation was then performed in Avizo using an algorithm based on distance mapping and thinning. The strut length distribution and average coordination number (CN) were computed using Avizo's built-in features. Bond and dihedral angles were computed in Matlab v.2021b from the positions of all connected nodes derived from the skeleton.

\section*{Data availability}
The data used in this manuscript is available at the Zenodo repository at dx.doi.org/10.5281/zenodo.7849558.


\begin{scriptsize}

\subsubsection*{Acknowledgements}

This study was financially supported by the Swiss National Science Foundation (SNSF) (163220, 188647, 168223, 190467), the National Center of Competence in Research \textit{Bio‐Inspired Materials} (51NF40-182881), and the Adolphe Merkle Foundation. This project had also received funding from the European Union's Horizon 2020 research and innovation programme under the Marie Sklodowska‐Curie grant agreement no. 706329/cOMPoSe (I.G.). This work was also funded under grant agreement no. 731019/EUSMI (J.L., D.K.) and made use of the Cornell Center for Materials Research Shared Facilities supported by the NSF MRSEC program (DMR-1719875). U.W. thanks the National Science Foundation (DMR‐1707836) for financial support. D.K. acknowledges funding from SNSF under grant no. 200021\_175905. J.L. and S.F. acknowledge support from the Japan Society for the Promotion of Science (JSPS) under KAKENHI 21K04816 and 19H05622, Cooperative Research Projects of CSIS, Tohoku University, and the Graduate Program for Spintronics (GP-Spin), Tohoku University. C.D. acknowledges support from the Max Planck Society Lise Meitner Excellence Program. The authors further acknowledge the Paul Scherrer Institut, Villigen, Switzerland for provision of synchrotron radiation beamtime at beamline X12SA (cSAXS) of the SLS.


\subsubsection*{Competing interests}
The authors declare no conflict of interest.

\end{scriptsize}

\printbibliography 

\newpage
\setcounter{figure}{0}
\renewcommand{\figurename}{Extended Data Fig.}
\onecolumn
\section*{Supporting Information Appendix (SI)}

\renewcommand\floatpagefraction{.001}
\makeatletter
\setlength\@fpsep{\textheight}
\makeatother

\begin{figure}[p!]
\centering
\includegraphics[width=\linewidth]{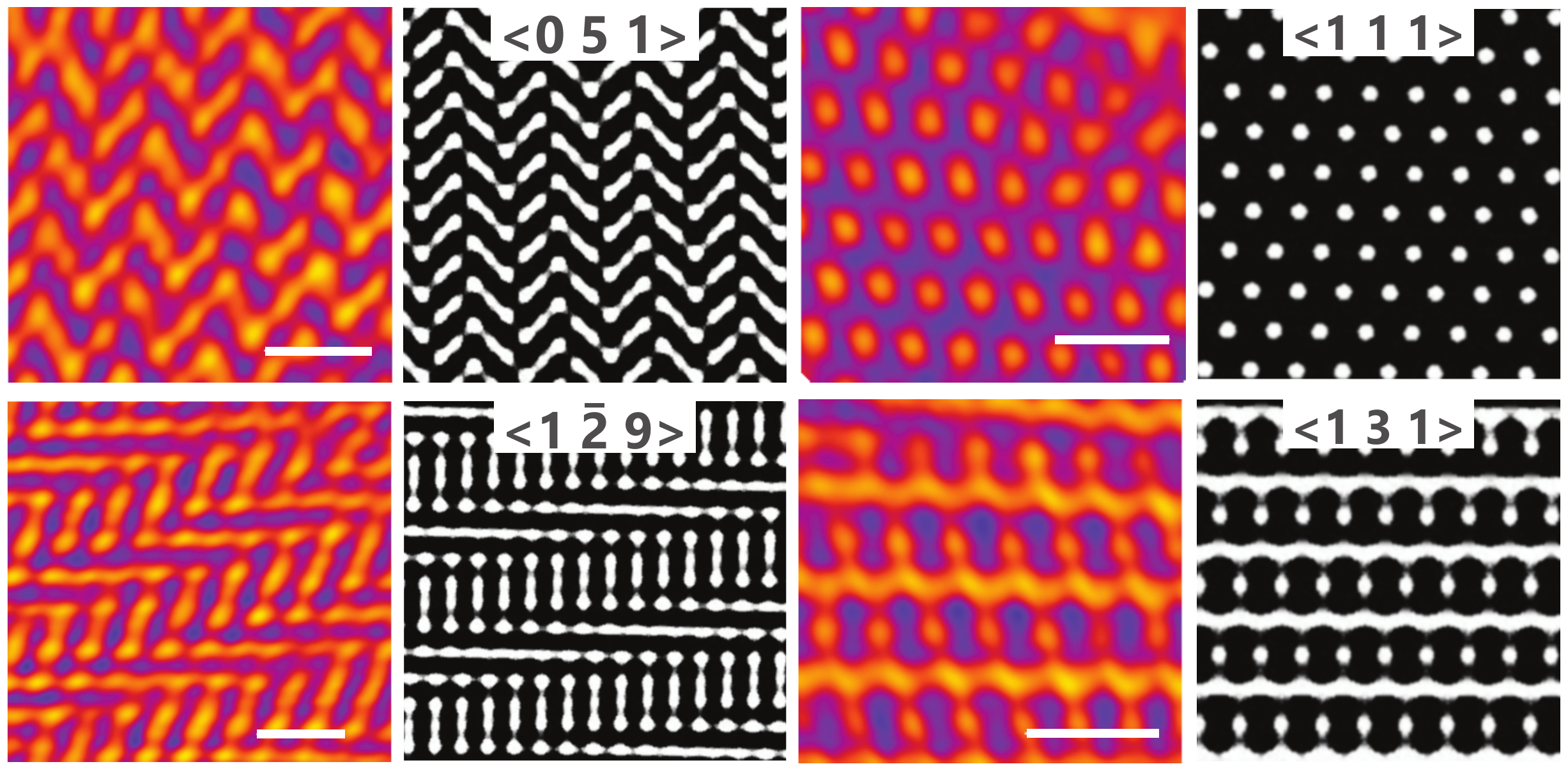}
\caption{\textbf{Additional experimental cross-sectional views.} Match between cross-sections extracted from the X-ray tomography dataset (orange-purple colourmap) and the $Fd\bar{3}m$ single diamond (SD) level-set (black and white). Miller indices of the SD level-set are written as $\langle\mathrm{hkl}\rangle$. Scale bars: 100\,nm.}
\label{figSI:add_CS}
\end{figure}

\newpage

\begin{figure}[!p]
\centering
\includegraphics[width=\linewidth]{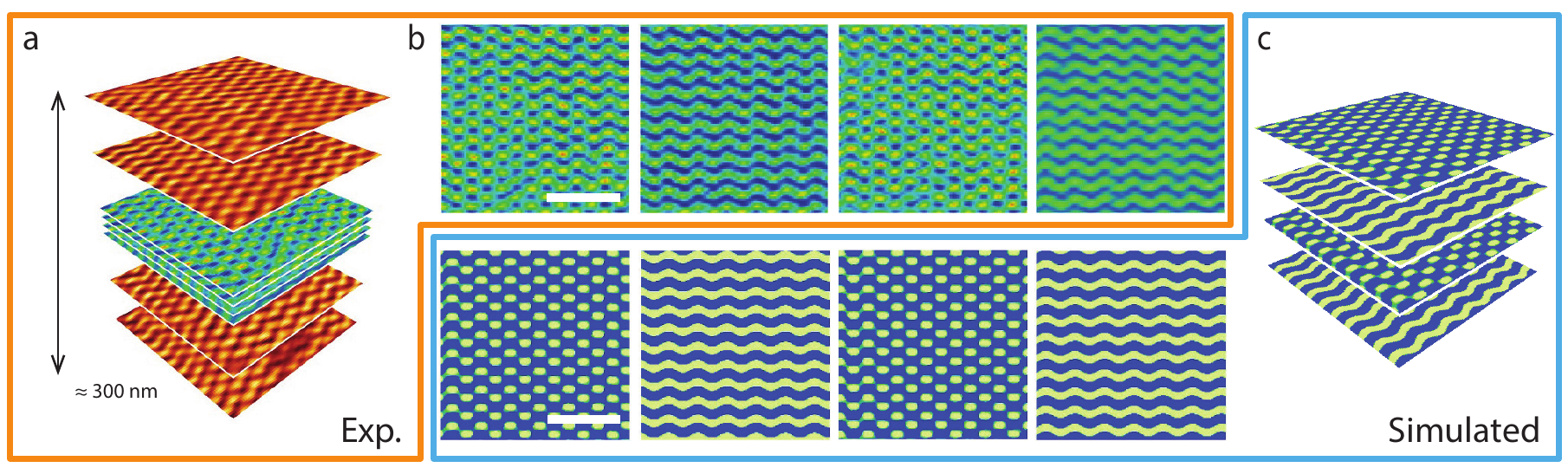}
\caption{\textbf{Experimental and modeled slices parallel to the substrate.} 
a) Experimental slices cut parallel to the substrate over a thickness of 300\,nm. The slices alternate between a wavy pattern and stretched hexagons throughout the film. 
b) Top row: experimental slices shown in a), viewed from top to bottom. The slices are spaced by 12\,nm. Bottom row: $\langle110\rangle$ oriented slices predicted by the distorted single diamond (DSD) model. Scale bars: 200\,nm. 
c) DSD $\langle110\rangle$ slices spaced by 13\,nm matching the patterns observed experimentally.}
\label{figSI:vertstack}
\end{figure}

\newpage

\begin{figure}[!t]
\centering
\includegraphics[width=0.95\linewidth]{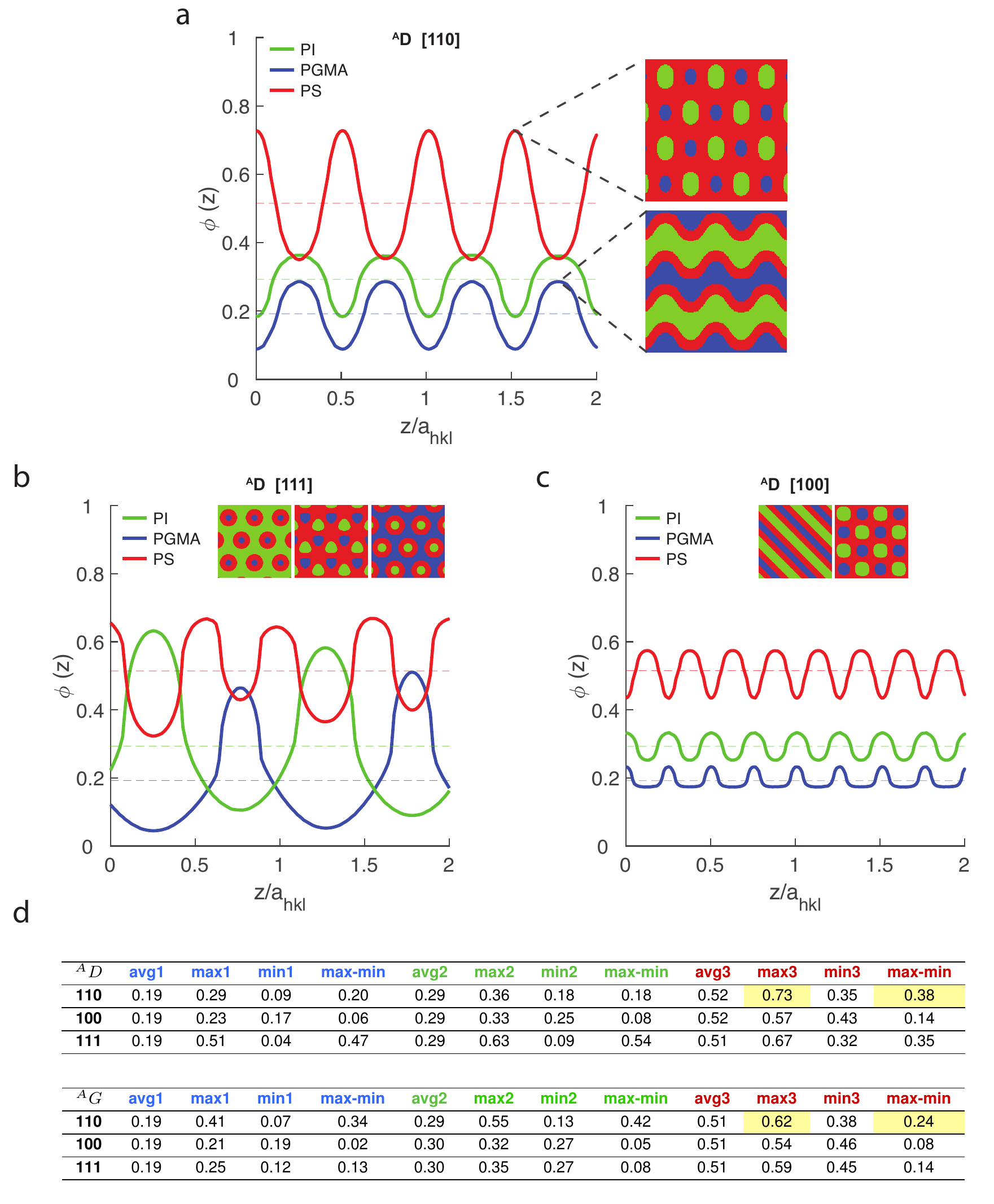} 
\caption{\textbf{Composition profile of a level-set alternating diamond.} Composition profile of each phase of the polymeric structure within two unit cells of an alternating diamond for the 
a) $\left[110\right]$, b) $\left[111\right]$, c) $\left[100\right]$ out-of-plane orientations. The fill fractions of the phases are $f_{\mathrm{PI}}=0.29$ shown in green, $f_{\mathrm{PGMA}}=0.19$ shown in blue, and $f_{\mathrm{PS}}=0.52$ shown in red. The two cross-sections shown represent the maximal and minimal area fraction of the matrix phase (corresponding to the minimal and maximal area fractions of the two other phases). The colour of each phase corresponds to the plot lines.
d) Area fraction of each phase for various orientations of an $^{A}D$ and an $^{A}G$. For each phase, the average value corresponds to the nominal volume fraction. The maximal and minimal value of the area fraction, as well as their difference, are calculated. The values shown here correspond to the plots in a--c.}
\label{figSI:ff_variation}
\end{figure}

\newpage

\begin{figure}[!p]
\centering
\includegraphics[width=\linewidth]{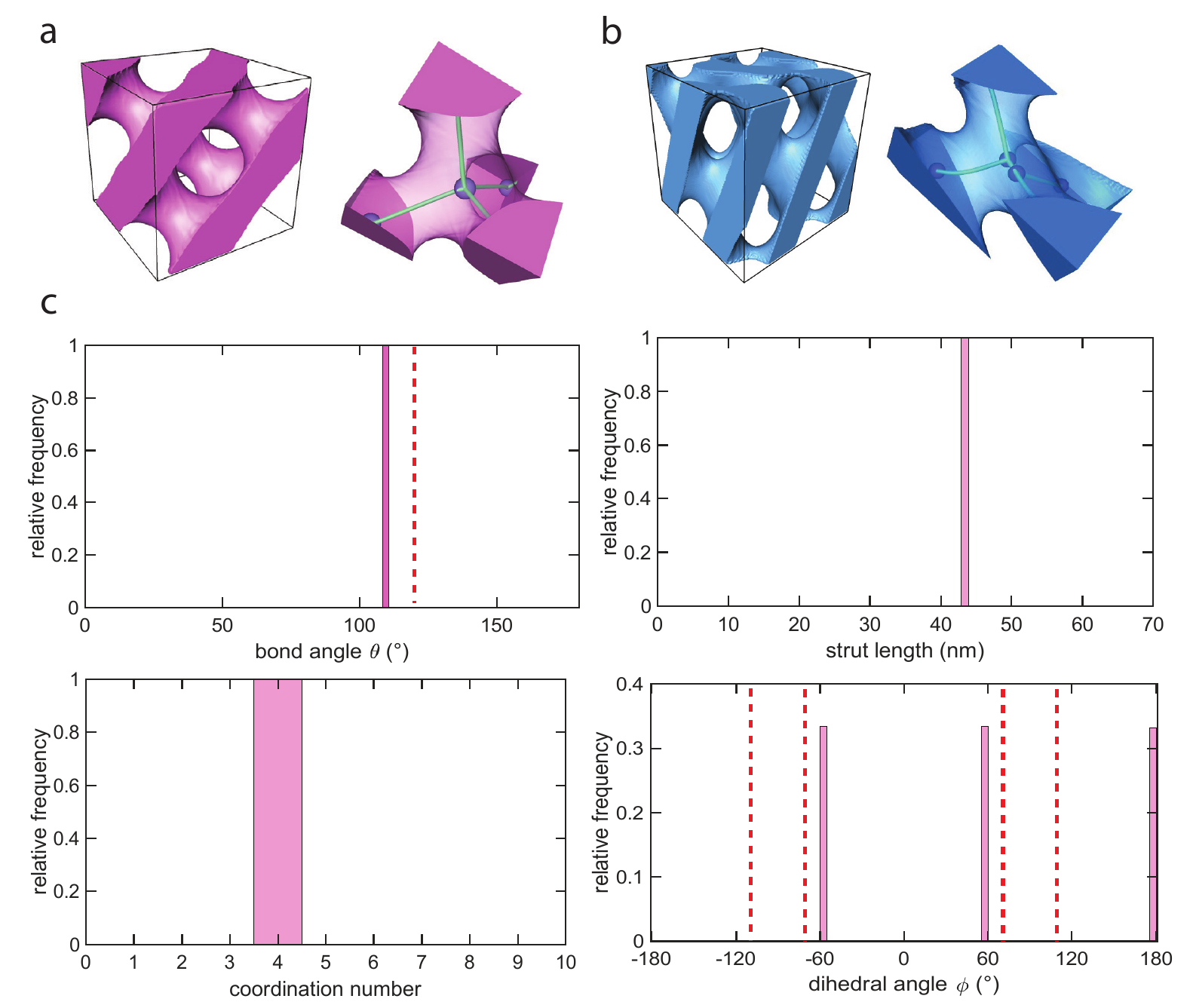}
\caption{\textbf{Cubic diamond and DSD.} Views of a) a level-set SD unit cell with $a=100$\,nm and $f=0.4$ and a single node compared with b) a DSD of same volume and its single strut. 
c) Statistical analysis of structural parameters of an SD: distributions of bond angle, strut length, coordination number and dihedral angles. The values expected for SD are $\theta=109.5$\textdegree\, , $\mathrm{CN}=4$, and $\varphi_d=\pm 60\degree, \pm 180\degree$. The values expected for a gyroid ($\theta=120$\textdegree\,, $\varphi_g=\pm 70.5\degree, \pm 109.5\degree$) are shown in red dotted lines as a reference.}
\label{figSI:diamond}
\end{figure}

\newpage

\begin{figure}[!p]
\centering
\includegraphics[width=\linewidth]{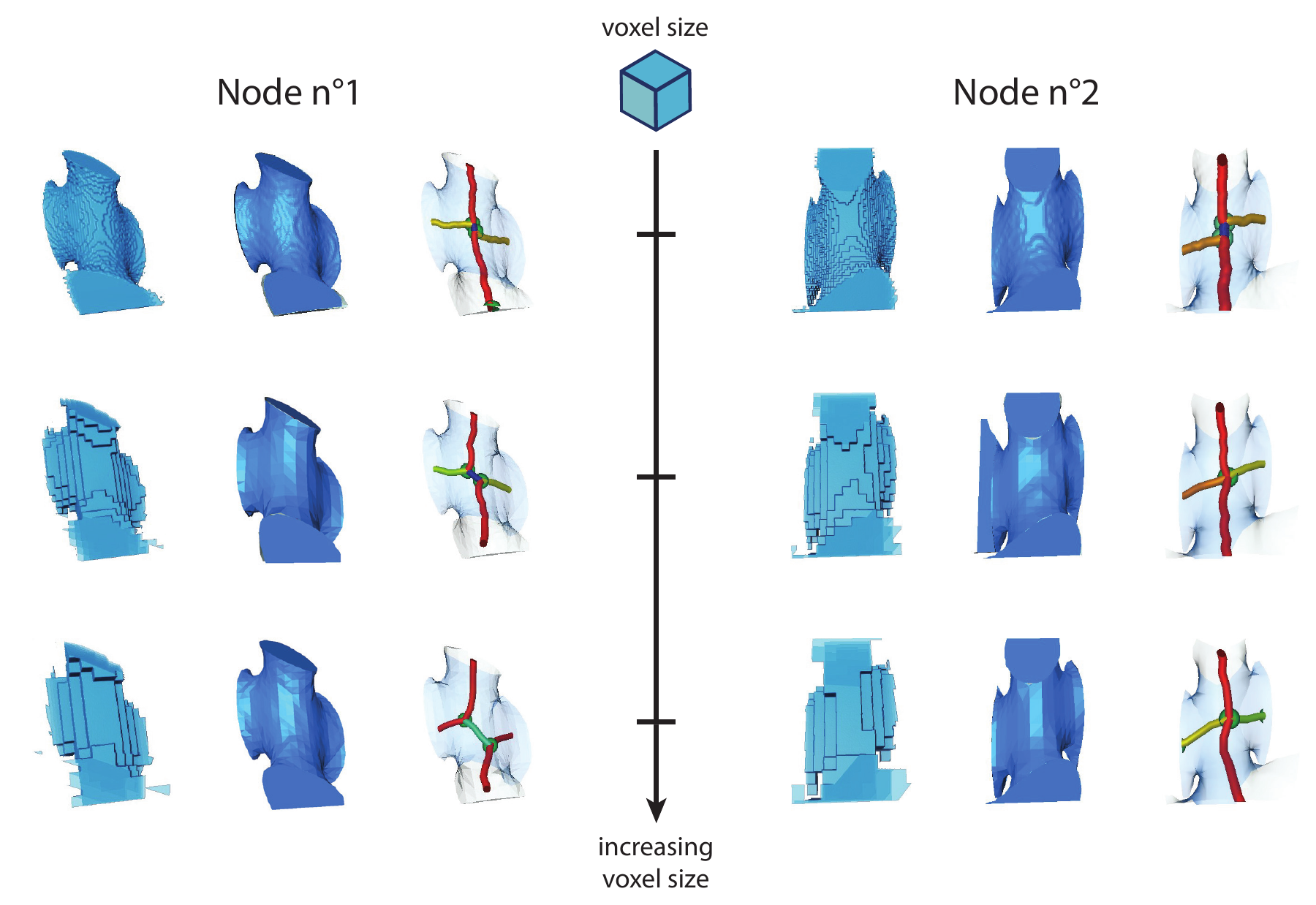}
\caption{\textbf{Voxel size-dependent skeletonisation in the DSD.} From left to right: voxelised volume rendering, smoothed surface rendering, skeleton. The strut colour corresponds to its length, with a blue to red colourmap corresponding to the $[5-35\,\mathrm{nm}]$ range. Upon decreasing the resolution (i.e. increasing the voxel size), some nodes do not change connectivity (Node n\textdegree1), while some have the two nodes linked by the shortest strut merge, thus changing the connectivity from 3 to 4 (Node n\textdegree2). This is caused by the dramatic loss of features evident in the voxelised rendering. Voxel size from top to bottom: 0.9\,nm, 2.6\,nm, 4.3\,nm.}
\label{figSI:vox_node}
\end{figure}

\newpage

\begin{figure}[!p]
\centering
\includegraphics[width=0.8\linewidth]{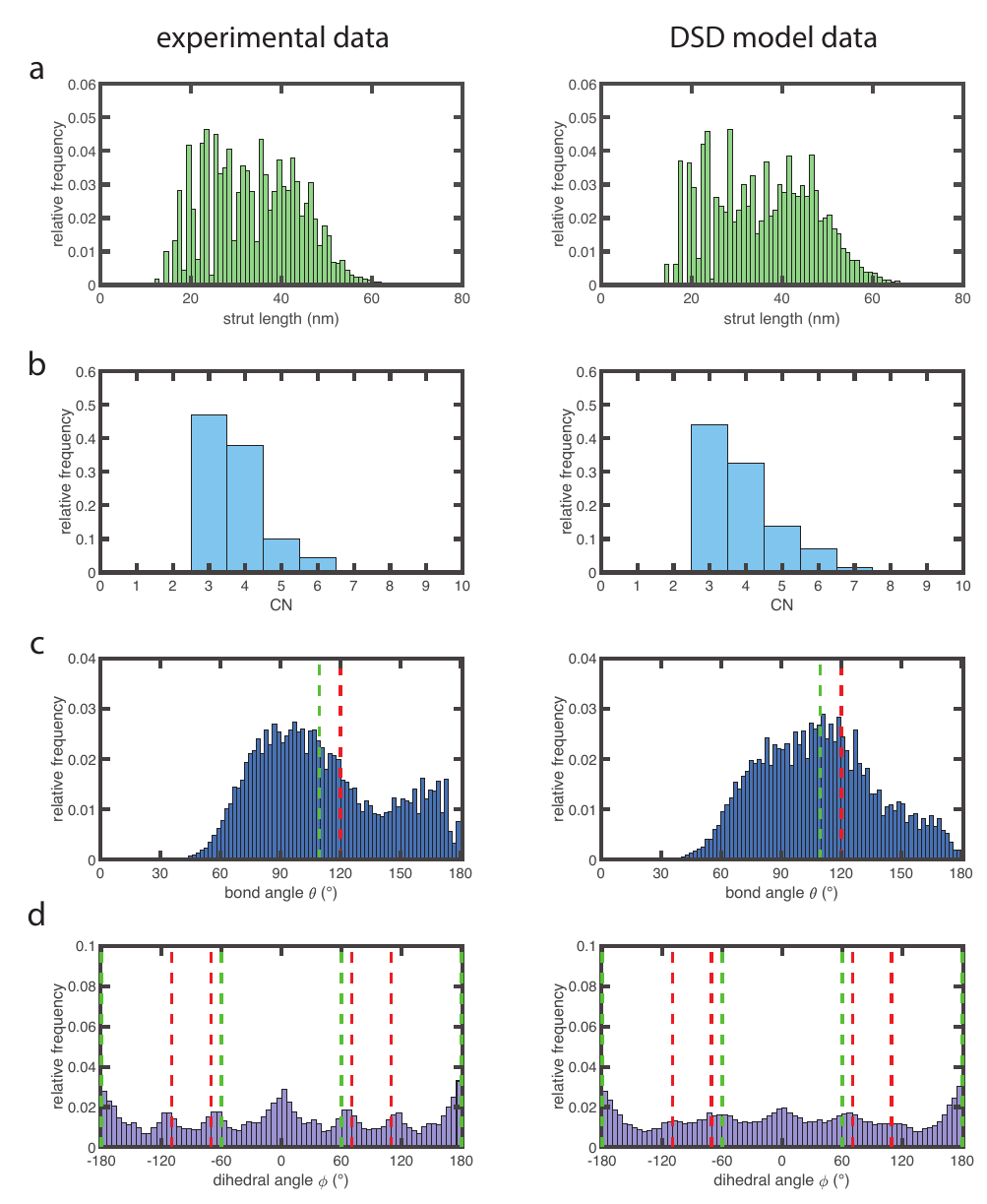}
\caption{\textbf{Comparison of structural parameters statistics calculated on the experimental dataset (left column) and the DSD model (right column)}. Distribution of the a) strut lengths ($N\approx20.000$), b) coordination numbers CN ($N\approx10.000$), c) bond angles $\theta$(\textdegree) ($N\approx60.000$), and d) dihedral angles $\phi$(\textdegree) ($N\approx100.000$). The volumes under investigation were $V=1250\times326\times1854\,$nm\textsuperscript{3}. The voxel size is 6.04\,nm in both cases. The angular values expected for a diamond and a gyroid network are shown in green and red dotted lines, respectively.} 
\label{figSI:vox_stat_ptycho}
\end{figure}

\newpage

\begin{figure}[!p]
\centering
\includegraphics[width=\linewidth]{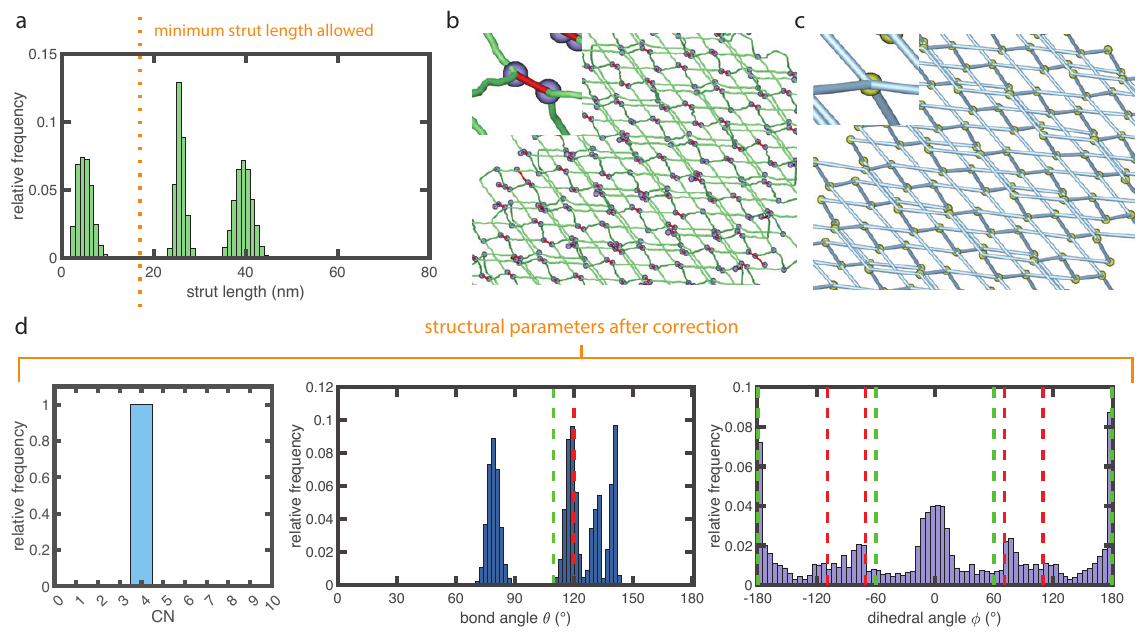}
\caption{\textbf{Structural parameters statistics after upsampling and correction of the skeleton.} 
a) A minimum strut length of 18\,nm is imposed so that two nodes connected by a smaller strut are merged into one ($N\approx16.000$). 
b) Experimental skeleton before correction: the struts smaller than 18\,nm are coloured in red. The majority of the nodes (purple spheres) appear to be doubled. Inset: magnification of a specific double node. 
c) Same skeleton after correction. Nodes that were doubled are now merged into a single (yellow) node. The general appearance of the skeleton is otherwise unchanged. Inset: same node as shown in b) after merger resulting from the correction. 
d) Structural parameters statistics from the corrected skeleton shown in c). The coordination number ($N\approx6.000$) is now equal to 4, as expected for a cubic diamond. The deviations of bond ($N\approx30.000$) and dihedral ($N\approx25.000$, computed on a smaller subvolume due to computational limitations) angles from the values expected for a cubic diamond (see Extended Data Fig.\,\ref{figSI:diamond}) reflect the distortions observed in the experimental sample. The values expected for the diamond and gyroid morphologies are shown as references in green and red, respectively. The voxel size is 1\,nm.}
\label{figSI:vox_stat_DSD}
\end{figure}

\newpage

\begin{figure}[!p]
\centering
\includegraphics[width=0.5\linewidth]{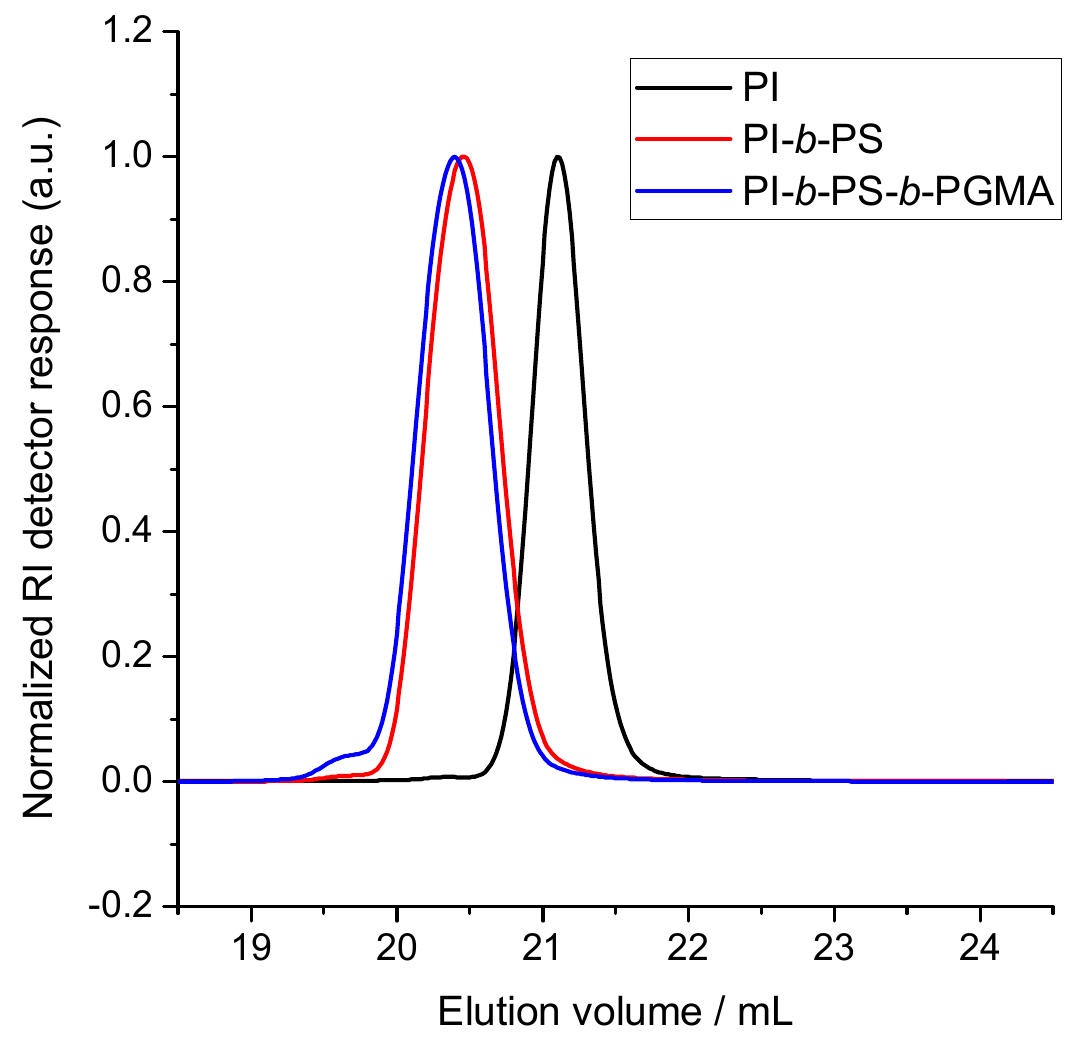}
\caption{\textbf{GPC} elugrams of PI homopolymer (black), PI-\textit{b}-PS diblock copolymer (red), and the final PI-\textit{b}-PS-\textit{b}-PGMA (ISG) triblock terpolymer (blue). The GPC was equipped with a Waters 410 refractive index detector and THF was used as the eluent. The elugrams were analyzed against a PI standard curve to deduce the number-averaged molar mass.}
\label{figSI:GPC}
\end{figure}

\newpage

\begin{figure}[!p]
\centering
\includegraphics[width=0.95\linewidth]{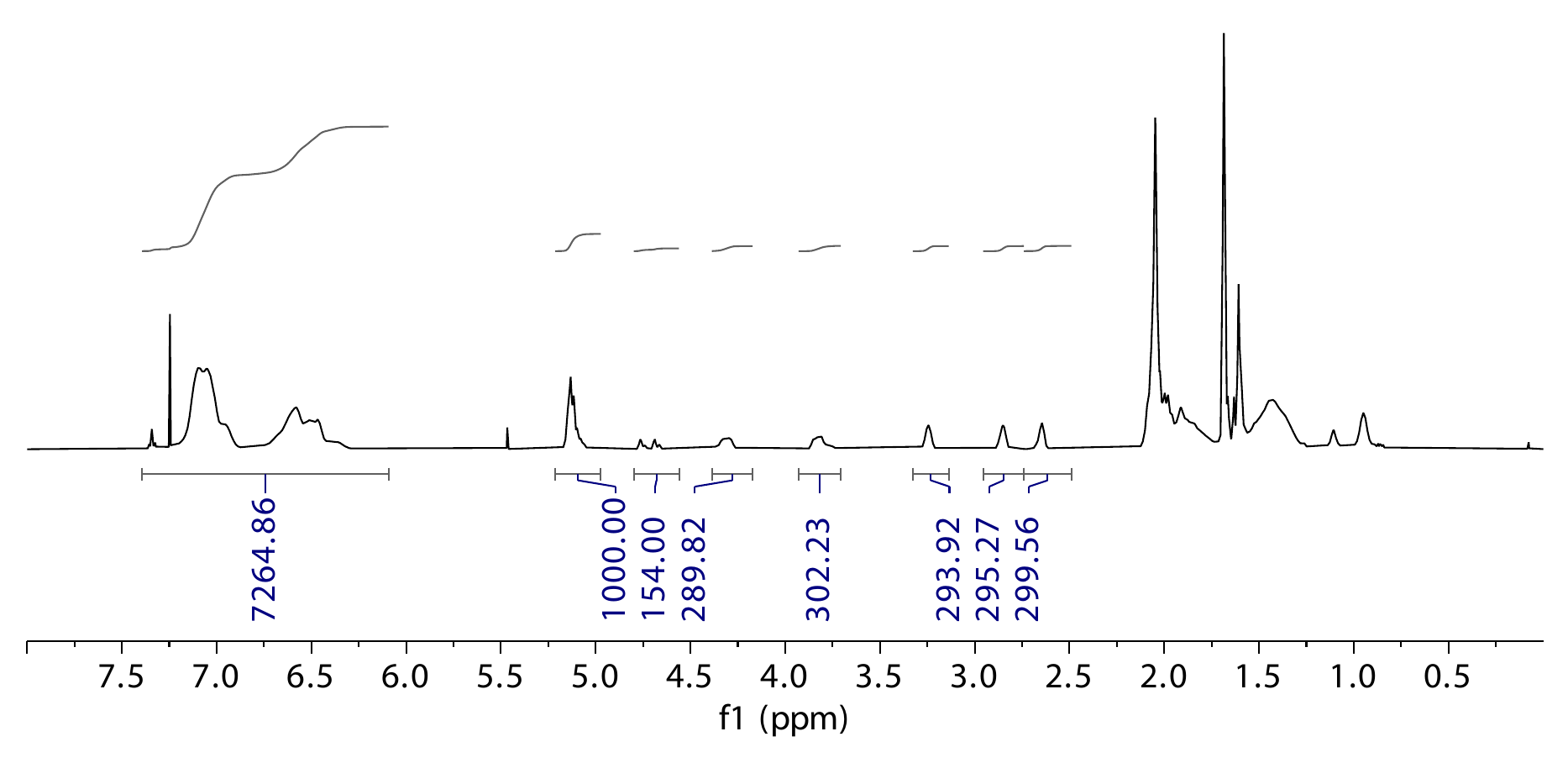}
\caption{\textbf{\textsuperscript{1}H NMR} spectrum of the ISG triblock terpolymer collected on a Varian Mercury-300 or Inova-400 spectrometer using chloroform-d as the solvent.}
\label{figSI:NMR}
\end{figure}

\newpage
\begin{figure}[!p]
\centering
\includegraphics[width=0.8\linewidth]{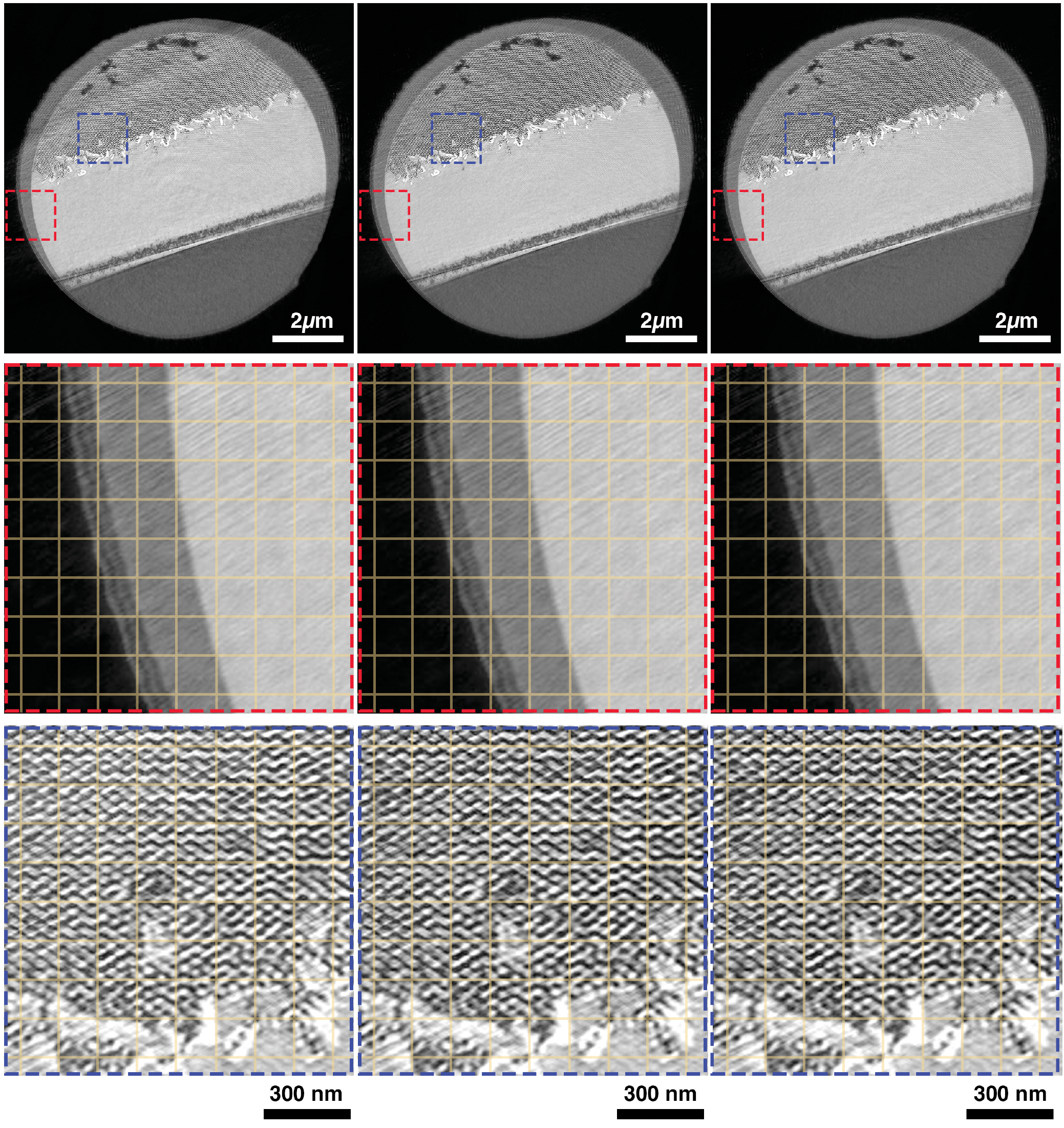}
\caption{\textbf{Sub-tomograms.} \textit{Top}: Slices through the 1st (left), 5th (center), and 8th (right) subtomogram. \textit{Middle}: Zoomed-in region marked with a red square in the corresponding top slice. \textit{Bottom}: Zoomed-in region marked with a blue square in the corresponding top slice. The semi-transparent grid is added to assist visual comparison of the structures.}
\label{figSI:PXCT}
\end{figure}

\end{document}